\documentclass[a4paper]{jpconf}

\usepackage[usenames]{xcolor}

\usepackage{amsfonts}
\usepackage{mathrsfs}
\usepackage{graphicx}
\usepackage{amsmath}
\usepackage{amssymb}
\usepackage{bm}
\usepackage{bbm}
\usepackage{epsfig}
\usepackage{multirow}
\usepackage{float}
\usepackage{array}
\usepackage{tabularx}

\definecolor{Code}{rgb}{0.444444,0.222222,0.}

\begin{document}

\graphicspath{{eps/}}

\title{Automated One-loop Computation in Quarkonium Process within NRQCD Framework}

\author{Feng Feng}
\address{Center for High Energy Physics, Peking University, Beijing 100871, China}
\ead{F.Feng@outlook.com}

\begin{abstract}
In last decades, it has been realized that the next-to-leading order corrections may become very important, and sometimes requisite, for some processes involving quarkoinum production or decay, e.g., $e^+e^- \to J/\psi + \eta_c$ and $J/\psi \to 3\gamma$. In this article, we review some basic steps to perform automated one-loop computations in quarkonium process within the Non-relativistic Quantum Chromodynamics~(NRQCD) factorization framework, and we give an introduction to some related public tools or packages and their usages in each step.
We start from generating Feynman diagrams and amplitudes with \textsc{FeynArts} for the quarkonium process, performing Dirac- and Color- algebras simplifications using \textsc{FeynCalc} and \textsc{FeynCalcFormLink}, and then to doing partial fractions on the linear-dependent propagators by \textsc{APart}, and finally to reducing the Tensor Integrals~(TI) into Scalar Integrals~(SI) or Master Integrals~(MI) using Integration-By-Parts~(IBP) method with the help of \textsc{Fire}. We will use a simple concrete example to demonstrate the basic usages of the corresponding packages or tools in each step.
\end{abstract}

\section{Generating Feynman Diagram and Amplitude}
Before starting to generate the corresponding Feynman diagram, we need to replace the incoming or outgoing hadronic states with corresponding partonic states, e.g., we need to replace $J/\psi$ with a quark and anti-quark pair $c(p_1)\bar{c}(p_2)$ for color-singlet case, and the pair $c\bar{c}$ with an extra gluon $c(p_1)\bar{c}(p_2) g(k)$ for the color-octet case. We take the process $(e^+e^-\to J/\psi+\eta_c)$ as an example, for the color-singlet model, the Feynman diagrams we actually want to generate are such process:
\begin{equation}
e^+ + e^- \to \gamma^* \to c(\frac{p_3}{2}+q_3) \bar{c}(\frac{p_3}{2}-q_3) + c(\frac{p_4}{2}+q_4) \bar{c}(\frac{p_4}{2}-q_4) \;.
\end{equation}

Currently, there are two tools or packages to automatically generate the Feynman diagrams:
\begin{itemize}
\item \textsc{FeynArts}\cite{Hahn:2000kx} is a Mathematica package for the generation and visualization of Feynman diagrams and amplitudes, it can be downloaded from \verb=http://www.feynarts.de/=.
\item \textsc{QGraf}\cite{Nogueira:1991ex} is a computer program with the programming language: \textsc{FORTRAN}~77,  that was written to assist in large perturbative calculations, in the context of Quantum Field Theory. It can generate Feynman diagrams and represent them by symbolic expressions, it can be downloaded from \verb=http://cfif.ist.utl.pt/~paulo/qgraf.html=.
\end{itemize}

Since we use \textsc{Mathematica} as our computation environment, we give a demonstration of the usage of \textsc{FeynArts} to generate the Feynman diagrams and amplitudes for the process:~$(e^+e^-\to\gamma^*\to J/\psi+\eta_c)$.
The basic steps are as follows:
\begin{itemize}
\item The underlying partonic process is:
$$\gamma^* \to c(\frac{p_3}{2}+q_3) \bar{c}(\frac{p_3}{2}-q_3) + c(\frac{p_4}{2}+q_4) \bar{c}(\frac{p_4}{2}-q_4) $$
\item Using {\tt CreateTopologies} to generate the topologies for the case $1\to4$:
{
\footnotesize
\begin{verbatim}
top=CreateTopologies[1,1->4,ExcludeTogologies->{WFCorrections, Tadpoles, V4onExt}];
\end{verbatim}
}
where we also exclude some topologies which don't contribute in our case.
\item Using {\tt InsertFields} to insert the fields in the corresponding model to the topologies we have just generated in the last step:
{
\footnotesize
\begin{verbatim}
tmp=InsertFields[top, {V[1]}->{F[3],-F[3],F[3],-F[3]}, Model->"SMQCD",
    ExcludeParticles->{V[1|2|3|4],S[_],F[4]}, InsertionLevel->{Clases}];
\end{verbatim}
}
\item Selecting or removing other unwanted diagrams using {\tt DiagramSelect} or {\tt DiagramDelete}, e.g.,
{
\footnotesize
\begin{verbatim}
all=DiagramDelete[tmp, 3...4, 13, 14, 25...25, 33...34, 42...43];
\end{verbatim}
}
\item Using {\tt CreateFeynAmp} to generate Feynman amplitudes for each diagram:
{
\footnotesize
\begin{verbatim}
amp=CreateFeynAmp[all, PreFactor->1]/.{FourMomentum[Incoming,1]->p3+p4,
    FourMomentum[Outgoing,1]->p3/2+q3, FourMomentum[Outgoing,2]->p3/2-q3,
    FourMomentum[Outgoing,3]->p4/2+q4, FourMomentum[Outgoing,4]->p4/2-q4,
    MQU[___]->mu, MQD[___]->md, EL->e, GS->GStrong};
\end{verbatim}
}
\item {\tt Export}ing the amplitudes to output file for later processing in \textsc{FeynCalc}\cite{Mertig:1990an}.
\end{itemize}

\section{Simplifying Dirac- and Color-Algebra}
To perform the Dirac- and Color-algebra simplifications, we adopt the covariant spin projectors techniques\cite{Bodwin:2002hg,Braaten:2002fi,Bodwin:2010fi} for the $q\bar{q}$ production:
\begin{eqnarray}
v(\bar{p}) \bar{u}(p) &\to& \frac{1}{4\sqrt{2}E(E+m_c)} (\bar{p}\!\!\!/-m_c)\big[\gamma_5,\epsilon\hspace{-.15cm}/^*\big](P\!\!\!\!/+2E)(p\!\!\!/+m_c)
\end{eqnarray}
where $\gamma_5$ and $\epsilon\hspace{-.15cm}/$ correspond to spin-singlet and spin-triplet respectively, and for $q\bar{q}$ decay, the projector reads:
\begin{eqnarray}
u(p) \bar{v}(\bar{p}) &\to& \frac{1}{4\sqrt{2}E(E+m_c)} (p\!\!\!/+m_c)(P\!\!\!\!/+2E) \big[\gamma_5,\epsilon\hspace{-.15cm}/\big] (\bar{p}\!\!\!/-m_c)
\end{eqnarray}
where $E$ is defined by:
\begin{equation}
E = \sqrt{m_c^2 + \left(\frac{p-\bar{p}}{2}\right)^2}
\end{equation}

Several packages or tools can be used to simplify the Dirac- and Color-algebra:
\begin{itemize}
\item {\sc Form}\cite{Kuipers:2012rf,Vermaseren:2000nd,Vermaseren:2011sb,Vermaseren:2010iw,Tentyukov:2008zz,Vermaseren:2008kw} is a Symbolic Manipulation System written in {\sc C} language, it can be downloaded from \verb=http://www.nikhef.nl/~form/=.
\item {\sc FeynCalc}\cite{Mertig:1990an} is a {\sc Mathematica} package for algebraic calculations in elementary particle physics, it can be downloaded from \verb=http://www.feyncalc.org/=.
\item {\sc FeynCalc/FormLink}\cite{Feng:2012tk} is developed to combine high-performance of {\sc Form} and user-friendliness of {\sc FeynCalc}, it can be downloaded from \verb=http://www.feyncalc.org/formlink/=.
\end{itemize}

Note that there is also another package {\sc FormCalc}\cite{hep-ph/9807565} which uses {\sc Form} from {\sc Mathematica}. The difference between {\sc FeynCalc/FormLink} and {\sc FormCalc} is the way in which {\sc Mathematica} and {\sc Form} communicate with each other. {\sc FormCalc} basically uses the method of input and output files, while {\sc FeynCalc/FormLink} uses the piping method. The basic idea of {\sc FormLink} is:
\begin{itemize}
\item {\sc FormLink} creates two unnamed pipes: {\tt r\#} and {\tt w\#}.
\item {\sc FormLink} starts {\sc Form} process with the command line: {\tt form -pipe r\#, w\# init}, where {\tt init} is just a initial {\sc Form} file with extension {\tt .frm}.
\item {\sc Form} sends its {\tt Process ID}({\tt PID}) to {\sc FormLink} in {\tt w\#}, and when {\sc FormLink} receives the {\tt PID}, it will responses two comma-separated {\tt PID} to {\sc Form}, the first one is the same as {\sc Form} {\tt PID}, and the last one corresponds to the {\tt PID} of {\sc FormLink}.
\item {\sc Form} start running the {\tt init.frm} file with the following codes: \par\vspace{.3cm}
\begin{tabular}{c|c}
\hline
\begin{minipage}{0.45\textwidth}
\begin{verbatim}

Off Statistics;
#ifndef ¡®PIPES_¡¯
#message "No pipes found";
.end;
#endif
#if (¡®PIPES_¡¯ <= 0)
#message "No pipes found";
.end;
#endif

\end{verbatim}
\end{minipage}
&
\begin{minipage}{0.45\textwidth}
\begin{verbatim}

#procedure put(fmt, mexp)
#toexternal ¡®fmt¡¯, ¡®mexp¡¯
#toexternal "#THE-END-MARK#"
#endprocedure
#setexternal ¡®PIPE1_¡¯;
#toexternal "OK"
#fromexternal
.end

\end{verbatim}
\end{minipage} \\
\hline
\end{tabular} \par\vspace{.3cm}
The key statement is {\tt \#fromexternal}, when {\sc Form} runs into this instruction, it will be blocked until the {\sc Form} code has been sent from {\sc Mathematica} through {\sc FormLink}, and then {\sc Form} will continue to execute the code which has been just sent.
\end{itemize}

Let us demonstrate the basic usage of {\sc FeynCalc/FormLink} with a simple example, i.e., the trace of six Dirac gamma matrix, first we calculate the trace with {\sc FeynCalc}:
\begin{verbatim}
  <<HighEnergyPhysics`fc`
  Tr[GS[p1, p2, p3, p4, p5, p6]]
\end{verbatim}
It is also quite simple to perform the trace with {\sc FeynCalcFormLink}, first we prepare the expression in {\sc FeynCalc} syntax, i.e.,
\begin{verbatim}
  exp = DiracTrace[GS[p1, p2, p3, p4, p5, p6]];
\end{verbatim}
note that we use {\tt DiracTrace} instead of {\tt Tr} to prevent the evaluation of the trace, then we just use {\tt FeynCalcFormLink} to calcuate the expression {\tt exp}:
\begin{verbatim}
  FeynCalcFormLink[exp]
\end{verbatim}
{\tt FeynCalcFormLink} first translate the {\tt exp} in {\sc FeynCalc} syntax to {\sc Form} code, for this simple case, the translated {\sc Form} code is as follows: \par\vspace{.1cm}
{\color{Code}
\noindent\rule{\textwidth}{0.3pt}\vspace{-0.2cm}
\begin{verbatim}
Vectors p1,p2,p3,p4,p5,p6;
Format Mathematica;
L resFL = (g_(1,p1)*g_(1,p2)*g_(1,p3)*g_(1,p4)*g_(1,p5)*g_(1,p6));
trace4,1;
contract 0;
.sort;
#call put("%E", resFL)
#fromexternal
\end{verbatim}\vspace{-0.3cm}
\rule{\textwidth}{0.3pt}
}
then the {\sc Form} code will be piped to {\sc Form} for execution, when {\sc Form} finishes running, it starts sending the result back to {\sc Mathematica}, and {\sc FeynCalcFormLink} will translate the result to {\sc FeynCalc} syntax.

{\tt SUNSimplify} in {\sc FeynCalc} can be used to perform the simplification on the color-algebra, e.g.,
\verb+SUNSimplify[SUNT[a,b,a,b]]+ to get the result
\begin{equation}
T^a T^b T^a T^b = -\frac{1}{2} C_F (C_A-2C_F).
\end{equation}
Before we are going to do the loop momentum integrals, we can use another technique, the method of region expansion\cite{Beneke:1997zp}, to greatly simplify our calculations. Usually, we expand the relative momentum $q$ between quark and anti-quark in quarkonium state after performing the loop integration, and then project the $S$-, $P$-, or $D$-waves. We can also expand the $q$ before the loop integration, as long as only the hard region is concerned according to the method  of region expansion\cite{Beneke:1997zp}. So if the NRQCD factorization is valid, it will be safe to use the method of region expansion to compute the short-distance coefficients, which correspond to the hard region.

\section{Passarino-Veltman Reduction for the Tensor Integrals}
The generic one-loop integral looks like:
\begin{equation}
\mathcal{T}^{\mu_1\cdots\mu_p} \equiv \frac{(2\pi\mu)^{4-d}}{i\pi^2} \int d^dk\frac{k^{\mu_1}\cdots k^{\mu_p}}{D_0D_1D_2\cdots D_{n-1}}
\end{equation}
where $D_i = (k+r_i) ^2 - m_i^2 + i \varepsilon$, $r_i = \sum_{k=1}^{i} p_k\, ( i=1\,,\,\cdots\,,\,n-1$), $r_0 = 0$ and $ r_{ij} = r_i - r_j $.
These one-loop integrals can be characterized by the so-called n-point tensor integrals, e.g.,
\begin{eqnarray}
B^{\mu}(r_{10}^2,m_0^2,m_1^20 &=&\frac{(2\pi\mu)^{4-d}}{i\pi^2} \int d^dk k^{\mu} \prod_{i=0}^1 \frac{1}{(k+r_i)^2-m_i^2}\\
C^{\mu\nu}(r_{10}^2,r_{12}^2,r_{20}^2,m_0^2,m_1^2,m_2^2) &=&\frac{(2\pi\mu)^{4-d}}{i\pi^2} \int d^dk k^{\mu}k^{\nu} \prod_{i=0}^2 \frac{1}{(k+r_i)^2-m_i^2} \nonumber\\
D^{\mu\nu\rho\sigma}(r_{10}^2,r_{12}^2,r_{23}^2,r_{30}^2,r_{20}^2,r_{13}^2,m_0^2,m_1^2,m_2^2,m_3^2) &=&\frac{(2\pi\mu)^{4-d}}{i\pi^2} \int d^dk  k^{\mu}k^\nu k^\rho k^\sigma \prod_{i=0}^3 \frac{1}{(k+r_i)^2-m_i^2} \nonumber
\end{eqnarray}
Generally, those n-point tensor integrals can be reduced to much simpler loop integrals, n-point scalar integrals:
\begin{eqnarray}
A_0(m_0^2) &=&\frac{(2\pi\mu)^{4-d}}{i\pi^2} \int d^dk \frac{1}{k^2-m_0^2} \\
B_0(r_{10}^2,m_0^2,m_1^2) &=&\frac{(2\pi\mu)^{4-d}}{i\pi^2} \int d^dk \prod_{i=0}^1 \frac{1}{(k+r_i)^2-m_i^2}  \nonumber\\
C_0(r_{10}^2,r_{12}^2,r_{20}^2,m_0^2,m_1^2,m_2^2) &=&\frac{(2\pi\mu)^{4-d}}{i\pi^2} \int d^dk \prod_{i=0}^2 \frac{1}{(k+r_i)^2-m_i^2}  \nonumber\\
D_0(r_{10}^2,r_{12}^2,r_{23}^2,r_{30}^2,r_{20}^2,r_{13}^2,m_0^2,m_1^2,m_2^2,m_3^2) &=&\frac{(2\pi\mu)^{4-d}}{i\pi^2} \int d^dk \prod_{i=0}^3 \frac{1}{(k+r_i)^2-m_i^2} \nonumber
\end{eqnarray}
We take a rank 4 tensor integral $D^{\mu\nu\rho\sigma}$ as an example, the tensor integral $D^{\mu\nu\rho\sigma}$ can be expressed as follows according to the Lorentz invariance:
\begin{eqnarray}
 D^{\mu\nu\rho\sigma}   & = & (g^{\mu\nu}g^{\rho\sigma}+g^{\mu\rho}g^{\nu\sigma}+g^{\mu\sigma}g^{\nu\rho})D_{0000} +\sum_{i,j,k,l=1}^{3} r_{i}^{\mu} r_{j}^{\nu} r_k^{\rho} r_l^{\sigma} D_{ijkl} \\
            && + \sum_{i,j}^{3} (g^{\mu\nu}r_{i}^{\rho} r_{j}^{\sigma}+g^{\nu\rho} r_{i}^{\mu} r_{j}^{\sigma}+g^{\mu\rho}r_i^{\nu}r_j^{\sigma}+g^{\mu\sigma}r_{i}^{\nu} r_{j}^{\rho}+g^{\nu\sigma} r_{i}^{\mu} r_{j}^{\rho}+g^{\rho\sigma} r_{i}^{\mu} r_{j}^{\nu})D_{00ij} \nonumber
\end{eqnarray}
where the $D_{0000}$, $D_{00ij}$ and $D_{ijkl}$ are some Lorentz scalar coefficients which can be expressed in terms of the n-point scalar integrals: $A_0$, $B_0$, $C_0$ and $D_0$. Such procedure is called Passarino-Veltman Reduction(PaVe-Reduction).

These coefficients can be achieved with the function {\tt PaVe} in {\sc FeynCalc}, e.g., the
\begin{center}
$D_{0000}(r_{10}^2,r_{12}^2,r_{23}^2,r_{30}^2,r_{20}^2,r_{13}^2,m_0^2,m_1^2,m_2^2,m_3^2)$
\end{center}
in {\sc FeynCalc} is expressed as:
\begin{center}
${\tt PaVe[0,0,0,0,},\{r_{10}^2,r_{12}^2,r_{23}^2,r_{30}^2,r_{20}^2,r_{13}^2\},\{m_0^2,m_1^2,m_2^2,m_3^2\}]$
\end{center}
where the first part in the argument of {\tt PaVe} function {\tt \{0,0,0,0\}} is the subscript of the corresponding $D_{0000}$ coefficient, and the remaining are the same as those in the argument of $D_{0000}$ coefficient. It should be noted that the factor involving renormalization scale $\mu$, i.e., $(2\pi\mu)^{4-d}$ has been dropped out in {\sc FeynCalc}.

Now to perform the PaVe-Reduction, we just use {\tt PaVeReduce},
\begin{verbatim}
  PaVeReduce[PaVe[0,0,0,0, {1,2,3,4,5,6}, {1,1,1,1}]]
\end{verbatim}
the output looks like
\begin{equation}
\begin{split}
-\frac{135 \text{C}_0(1,2,5,1,1,1)}{2401}-\frac{5751 \text{C}_0(1,4,6,1,1,1)}{192080}-\frac{88691\text{C}_0(2,3,6,1,1,1)}{2650704} \\
-\frac{5755 \text{C}_0(3,4,5,1,1,1)}{633864}+\frac{1587 \text{D}_0(1,2,3,4,5,6,1,1,1,1)}{38416}+\frac{51 \text{B}_0(1,1,1)}{1960} \\
+\frac{907 \text{B}_0(2,1,1)}{27048}+\frac{1025 \text{B}_0(3,1,1)}{99176}+\frac{347 \text{B}_0(4,1,1)}{32340}
-\frac{50 \text{B}_0(5,1,1)}{1617}-\frac{181 \text{B}_0(6,1,1)}{22540}+\frac{5}{72}
\end{split}
\end{equation}
where we take some special numerical values for the argument of $D_{0000}$. We can see that the coefficient $D_{0000}$ is now expressed in terms of n-point scalar integrals: $B_0$, $C_0$ and $D_0$.

Since the PaVe-Reduction is based on solving linear equations, generally it will encounter some problems when the Gram determinant equals $0$, which happens if we expand the relative momentum $q$ before loop integration, due to taking the derivative over $q$. So we need some more general method called Integration-By-Parts (IBP) reduction to perform the reduction of tensor integrals.

\section{Partial Fraction and IBP Reduction}
Let us consider a general Feynman integral, here we adopt the notation as in \cite{Smirnov:2008iw},
\begin{equation}
F(a_1,\cdots,a_n) = \int\cdots\int\frac{d^dk_1\cdots d^dk_h}{E_1^{a_1}\cdots E_n^{a_n}}
\end{equation}
where $k_i$, $i = 1, \cdots , h$, are loop momenta and the denominators $E_r$, $r=1,\cdots,n$, are either quadratic or linear with respect to the loop momenta $k_i$ of the graph. Irreducible polynomials in the numerator can be represented as denominators raised to negative powers.

The basic idea of IBP reduction is that, we know the integration of such derivative is $0$, i.e.,
\begin{equation}
\int\cdots\int d^d k_1 d^d k_2 \cdots \frac{\partial}{\partial k_i} \left[\frac{p_j}{E_1^{a_1}\cdots E_n^{a_n}}\right] = 0
\end{equation}
where $k_i$ are the loop momenta, and $p_j$ are the momenta which can internal or external, so with different $k_i$ and $p_j$, we can get a list of equations which can be expressed as follows:
\begin{equation}
\sum \alpha_i F(a_1+b_{i,1},\cdots,a_n+b_{i,n}) = 0
\end{equation}
By solving these equations, we can express the complicated loop integrals in terms of much simpler ones, which we call Master Integral~(MI).

There are many packages or tools in the market which can be used to perform the IBP reduction, e.g.,
\begin{itemize}
\item {\sc AIR}\cite{Anastasiou:2004vj} is {\sc Maple} package, which can be downloaded from \\ \verb=http://www.phys.ethz.ch/~pheno/air/=.
\item {\sc Fire}\cite{Smirnov:2008iw} is {\sc Mathematica} package, which can be downloaded from \\ \verb=http://science.sander.su/FIRE.htm=.
\item {\sc Reduze}\cite{Studerus:2009ye} is written in {\tt C}, which can be downloaded from \\ \verb=http://reduze.hepforge.org/=.
\item {\sc LiteRed}\cite{Lee:2012cn} is another {\sc Mathematica} package, which can be downloaded from \\ \verb=http://www.inp.nsk.su/~lee/programs/LiteRed/=.
\item Many other private codes.
\end{itemize}

There is a precondition to perform the IBP reduction, i.e., the propagators should be linear independent, so we need another {\sc Mathematica} package {\sc APart}\cite{Feng:2012iq} to perform partial fraction on the propagators.

Let us take a simple physical loop integral to demonstrate the usage of {\sc APart} and {\sc Fire},
\begin{equation}
{\tt exp} = \frac{(k\cdot p_1) \, (k\cdot p_2)} {k^2[(k+p_1)^2-m^2][(k+p_2)-m^2]^2}
\end{equation}
the linear independent variables involving loop momentum are $k^2,\,k\cdot p_1,\,k\cdot p_2$, which can be expressed in {\sc FeynCalc}:
\begin{verbatim}
  xs = FCI/@{ SP[k], SP[k, p1], SP[k, p2] }
\end{verbatim}
then to perform partial fraction on the loop integral is ready with {\sc APart}:
\begin{verbatim}
  $APart[exp, xs]
\end{verbatim}
the result looks like:
{\scriptsize
\begin{eqnarray}\label{RedBox}
&&\frac{1}{4} \left(m^2-\text{p1}^2\right) \left\| \frac{1}{\left(-k^2+m^2-\text{p1}^2-2 k\cdot \text{p1}\right) \left(-k^2+m^2-\text{p2}^2-2 k\cdot \text{p2}\right)^2}\right\| \\
&&+\frac{1}{4} \left(m^2-\text{p1}^2\right) \left\| \frac{1}{k^2 \left(k^2-m^2+\text{p1}^2+2 k\cdot \text{p1}\right) \left(k^2-m^2+\text{p2}^2+2 k\cdot \text{p2}\right)}\right\| \nonumber\\
&&+\frac{1}{2}\; {\color{red}\fbox{\color{black}$\displaystyle\left\| \frac{k\cdot \text{p2}}{\left(-k^2+m^2-\text{p1}^2-2 k\cdot \text{p1}\right) \left(k^2-m^2+\text{p2}^2+2 k\cdot \text{p2}\right)^2}\right\|$}} \nonumber\\
&&+\frac{1}{4} \left(m^2-\text{p1}^2\right) \left(m^2-\text{p2}^2\right) \left\| \frac{1}{k^2 \left(k^2-m^2+\text{p1}^2+2 k\cdot \text{p1}\right) \left(k^2-m^2+\text{p2}^2+2 k\cdot \text{p2}\right)^2}\right\| \nonumber\\
&&-\frac{1}{4} \left\| \frac{1}{\left(-k^2+m^2-\text{p2}^2-2 k\cdot \text{p2}\right)^2}\right\| +\frac{1}{4} \left\| \frac{1}{k^2 \left(k^2-m^2+\text{p2}^2+2 k\cdot \text{p2}\right)}\right\| +\frac{1}{4} \left(m^2-\text{p2}^2\right) \left\| \frac{1}{k^2 \left(k^2-m^2+\text{p2}^2+2 k\cdot \text{p2}\right)^2}\right\| \nonumber
\end{eqnarray}
}
we can see that there are at most three propagators in each term, and these propagators in each term are liner independent now.

Finally we can use {\sc Fire} to perform IBP reduction, we take the tensor integral framed with red box in Eq.~(\ref{RedBox}) as an example, such integral can be expressed as {\tt F[\{-1,1,2\}]} with {\tt F} defined by:
\begin{equation}\label{FDef}
\mbox{\tt F[\{l, m, n\}]} = \int\frac{d^4 k}{(2\pi)^4}\frac{(k\cdot p_2)^{-l}}{(m^2-k^2-2k\cdot p_1-p_1^2)^m(-m^2+k^2+2k\cdot p_2+p_2^2)^n}
\end{equation}
The basic usage of {\sc Fire} is like this:
{\small
\begin{verbatim}
Replacement = {p1^2 -> m^2, p2^2 -> m^2, p1 p2 -> SP[p1, p2]};
Internal = {k};
External = {p1, p2};
Propagators = {k p2, -2 k p1 - k^2 + m^2 - p1^2, 2 k p2 + k^2 - m^2 + p2^2};
PrepareIBP[];
startinglist = {IBP[k, k], IBP[k, p1], IBP[k, p2]}/.Replacement;
Prepare[];
Burn[];
\end{verbatim}
}
\noindent first we input the internal and external momenta, and provide the independent propagators in {\tt Propagators}, then prepare the IBP equations with {\tt startinglist}, finally {\tt Burn} in {\sc Fire}, and now it is ready to get the result for {\tt F[\{-1, 1, 2\}]}, just use the {\tt F} function:
\begin{equation}
\mbox{\tt F[\{-1, 1, 2\}]} = \frac{(d-2) G(\{0,0,1\})}{8 \left(m^2-\text{p1}\cdot \text{p2}\right)}+\frac{(d-2) G(\{0,1,0\})}{8 \left(m^2-\text{p1}\cdot \text{p2}\right)}+\frac{1}{4} (4-d) G(\{0,1,1\})
\end{equation}
where the definition $G$ is the same as {\tt F} in Eq.~(\ref{FDef}).

We can apply such procedure to each loop integral in Eq.~(\ref{RedBox}) to get the finally IBP reduced result:
\begin{eqnarray}
&&\left\|\frac{(k\cdot p_1) \, (k\cdot p_2)} {k^2[(k+p_1)^2-m^2][(k+p_2)-m^2]^2}\right\| \nonumber\\
&\Rightarrow&\frac{(D-2) \left\| \frac{1}{-k^2-2 k\cdot \text{p1}}\right\| }{16 \left(m^2-\text{p1}\cdot \text{p2}\right)}+\frac{(D-2) \left\| \frac{1}{k^2+2 k\cdot \text{p2}}\right\| }{16 \left(m^2-\text{p1}\cdot \text{p2}\right)} \nonumber\\
&&+\frac{1}{8} (4-D) \left\| \frac{1}{\left(-k^2-2 k\cdot \text{p1}\right) \left(k^2+2 k\cdot \text{p2}\right)}\right\| +\frac{1}{4} \left\| \frac{1}{k^2 \left(k^2+2 k\cdot \text{p2}\right)}\right\|
\end{eqnarray}
where we set $p_1^2=p_2^2=m^2$ to simplify the result, and $\left\|\cdots\right\|$ is defined by
\begin{equation}
\left\| {\rm exp} \right\| = \int\frac{d^D k}{(2\pi)^D} \; {\tt exp}
\end{equation}
It can be seen that the original tensor integral has been reduced to much simper scalar integrals or master integrals, and we can apply such procedure to each tensor integral in each Feynman diagram, and get the final expression expressed in terms of scalar integrals, which can be calculated, analytically or numerically, by any other means.
\section*{Acknowledgments}

The author would like to thank Rolf Mertig for the collaboration in the {\sc FormLink} and {\sc FormLinkFeynCalc} project, and to Prof. Yu Jia, Postdoc Wen-Long Sang and Hai-Rong Dong for many useful discussions in the related works.

Finally, Feng Feng would like to commemorate his beloved mother and father.


\end{document}